\documentclass[conference,10pt]{IEEEtran}
\IEEEoverridecommandlockouts
\usepackage{cite}
\usepackage{amsmath,amssymb,amsfonts}
\usepackage{algorithmic}
\usepackage{graphicx}
\usepackage{textcomp}
\usepackage{xcolor}
\usepackage{subfigure}
\usepackage{makecell}
\usepackage{tabularx}
\usepackage{multirow}
\usepackage{todonotes}
\usepackage{tcolorbox}
\usepackage{diagbox}
\usepackage{array}
\usepackage{verbatim}
\usepackage{caption}

\makeatletter
\renewcommand{\citepunct}{,\penalty\@m\hskip.13emplus.1emminus.1em}
\renewcommand{\citedash}{\hbox{--}\penalty\@m}

\def\BibTeX{{\rm B\kern-.05em{\sc i\kern-.025em b}\kern-.08em
    T\kern-.1667em\lower.7ex\hbox{E}\kern-.125emX}}

\begin{document}

\title{Precoder Learning by Leveraging Unitary Equivariance Property
}

\author{
\IEEEauthorblockN{Yilun Ge, Shuyao Liao, Shengqian Han, Chenyang Yang}
\IEEEauthorblockA{School of Electronics and Information Engineering, Beihang University, Beijing 100191, China\\Email: \{yilunge, shuyaoliao, sqhan, cyyang\}@buaa.edu.cn}

}

\maketitle

\begin{abstract}
Incorporating mathematical properties of a wireless policy to be learned into the design of deep neural networks (DNNs)
is effective for enhancing learning efficiency. Multi-user precoding policy in multi-antenna system, which is the mapping from channel matrix to precoding matrix, possesses a permutation equivariance property, which has been harnessed to design the parameter sharing structure of the weight matrix of DNNs. In this paper, we study a stronger property than permutation equivariance, namely unitary equivariance, for precoder learning. We first show that a DNN with unitary equivariance designed
by further introducing parameter sharing into a permutation equivariant DNN is unable to learn the optimal precoder. We proceed to develop a novel non-linear weighting process satisfying unitary equivariance and then construct a joint unitary and permutation equivariant DNN. Simulation results demonstrate that the proposed DNN not only outperforms existing learning methods in learning performance and generalizability but also reduces training complexity.
\end{abstract}
\section{Introduction}
\label{sec:intro}

The optimization of multi-user multi-input multi-output (MU-MIMO) precoder is a challenging problem. Various numerical algorithms have been developed, e.g., the weighted minimum mean square error (WMMSE) algorithm~\cite{shi2011iteratively}, which, however, are with high computational complexity.

Deep learning offers a promising solution for optimization due to its low inference complexity. In~\cite{sun2018learning}, a fully-connected deep neural network (DNN) was trained to approximate the performance of WMMSE with significantly reduced inference complexity. However, such DNNs often involve a large number of learnable parameters, requiring extensive training time and massive training samples.
Incorporating inductive biases into DNNs architecture with mathematical properties of the policies being learned is crucial for enhancing learning efficiency~\cite{ravanbakhsh2017equivariance,lim2022enn}. A well-known example is convolutional neural network (CNN), which leverages the property of translational invariance and has significantly promoted the field of image~recognition~\cite{lecun1989backpropagation}.

Recent studies have demonstrated the importance of integrating permutation equivariance properties to enhance learning efficiency for wireless communications~\cite{Eisen2020,He2021,zhao2022learning,10108002,guo2023model,multidimensional,10333362}.  In~\cite{zhao2022learning}, the mismatch between the inductive bias of CNNs and the permutation equivariance property of the precoding policy was identified, and an edge-updated graph neural network (GNN) was proposed to leverage the permutation equivariance property. In~\cite{10108002}, the attention mechanism was introduced into GNN and a graph attention network (GAT) was used to learn the energy efficient precoder.
In \cite{guo2023model}, a GNN was designed by introduce Taylor's expansion of matrix pseudo-inverse into the model, which improves learning efficiency. In \cite{multidimensional}, a multidimensional GNN was developed for hybrid precoder learning, which is able to exploit complicated permutation equivariance properties.

In this paper, we study the learning of MU-MIMO precoder by investigating a stronger property than permutation equivariance, known as unitary equivariance. This property means that if the channel matrix is multiplied by a unitary matrix, then the optimal precoder will be multiplied by the same unitary matrix.
The only instance of a unitary equivariant DNN design was introduced in \cite{ma2023feedforward}. However, it lacked a systematic analysis of how the unitary equivariance property impacts DNN design, and its performance in precoder learning is unsatisfactory, as to be demonstrated later.
In this paper, we first show that a DNN with a parameter sharing structure derived from the unitary and permutation equivariance property is unable to learn the optimal precoder. To solve the problem, we then develop a non-linear weighting process that satisfies the unitary equivariance, and finally construct a joint unitary and permutation equivariant neural network (UPNN) for precoder learning. Simulation results demonstrate the advantages of the proposed UPNN in performance improvement and training complexity reduction.


\section{Equivariance Property of a Precoding Policy}
\label{sec:precode}

Consider a downlink MU-MIMO system where a base station (BS) equipped with $ N $ antennas serves $ K $ single-antenna users. The channel matrix is denoted as $\mathbf{H}=[\mathbf{h}_1, \dots, \mathbf{h}_K] \in \mathbb{C}^{N\times K}$, where $ \mathbf{h}_k\in \mathbb{C} ^{N\times 1} $ is the channel vector of user $k$. The precoding matrix is denoted as $ \mathbf{V}=[ \mathbf{v}_1, \dots, \mathbf{v}_K] \in \mathbb{C}^{N\times K}$, where $ \mathbf{v}_k\in \mathbb{C} ^{N\times 1} $ is the precoding vector of user $k$. The precoder optimization problem that maximizes the sum rate subject to the total power constraint is
\begin{subequations}\label{E:precodModel}
	\begin{align}
		\max_{\mathbf{V}} \quad & \sum\nolimits_{k=1}^K{R_k} \label{E:precodModela}\\
		\rm{s.t.}\quad
		&\mathrm{Tr}\left( \mathbf{V}^{\mathrm{H}}\mathbf{V} \right) \le P_{\text{max}}, \label{E:precodModelb}
	\end{align}
\end{subequations}
where $R_k=\log _2\Big( 1+\frac{\left| \mathbf{h}_{k}^{\mathrm{H}}\mathbf{v}_k \right|^2}{\sum_{m\ne k}{\left| \mathbf{h}_{k}^{\mathrm{H}}\mathbf{v}_m \right|^2}+\delta _{k}^{2}} \Big)$ is the data rate of user $k$, $ \delta _{k}^{2} $ is the noise power, and $P_{\text{max}}$ is the maximal transmit power of the BS.

The precoding policy is the mapping from a channel matrix $\mathbf{H}$ to its corresponding optimal precoding matrix $\mathbf{V}^*$, i.e., $ \mathbf{V}^*=\mathcal{F}\left( \mathbf{H} \right)$. When $\mathbf{H}$ is transformed by left multiplying with a unitary matrix $\mathbf{U}$ and right multiplying with a permutation matrix $\mathbf{\Pi}^{\text{T}}$, it follows from problem \eqref{E:precodModel} that the correspondingly transformed $\mathbf{V}^*$ remains an optimal solution. Hence, the precoding policy $\mathcal{F}( \cdot)$ exhibits a joint unitary and permutation equivariance (UE-PE) property, which can be expressed as
\begin{equation}\label{E:18}
	\mathbf{U{V}^*\Pi }^{\text{T}}=\mathcal{F}\left( \mathbf{UH\Pi }^{\text{T}} \right),
\end{equation}
where $\mathbf{U}\in\mathbb{C}^{N\times N}$ and $\mathbf{\Pi}\in\mathbb{R}^{K\times K}$ are arbitrary unitary and permutation matrix, respectively.

A permutation matrix is a special type of unitary matrix. Thus, the equation in \eqref{E:18} remains valid when replacing $\mathbf{U}$ with another permutation matrix $\widetilde{\mathbf{\Pi}}$. Consequently, the UE-PE property can be reduced to the two-dimensional PE (2D-PE) property, as defined in~\cite{zhao2022learning}.

\section{DNNs with the PE Property}
\label{sec:pe_dnns}
When using a DNN to find the precoding policy from the non-convex problem in \eqref{E:precodModel}, for notational simplicity, we use $\mathbf{D}^{\prime}\in \mathbb{C}^{N\times K\times M^{\prime}}$ and $\mathbf{D}\in \mathbb{C}^{N\times K\times M}$ to denote the output and input of a hidden layer of the DNN, which have $M^{\prime}$ and $M$ hidden representations, respectively. Here, we omit the layer number in $\mathbf{D}^{\prime}$ and $\mathbf{D}$. For simplicity, we set $M^{\prime}$ and $M$ to one, which can be extended to any positive integer easily. With $M^{\prime}=M=1$, $\mathbf{D}^{\prime}$ and $\mathbf{D}$ reduce to two matrices in $\mathbb{C}^{N\times K}$, which can be expressed as $\mathbf{D}^{\prime}=[\mathbf{d}^{\prime}_{1}, \dots, \mathbf{d}^{\prime}_{K}]$ and $\mathbf{D}=[\mathbf{d}_{1}, \dots, \mathbf{d}_{K}]$, where $\mathbf{d}^{\prime}_{k}=[d^{\prime}_{1,k}, \dots, d^{\prime}_{N,k}] ^{\text{T}}\in \mathbb{C}^{N\times 1}$ and $\mathbf{d}_{k}=[d_{1,k}, \dots, d_{N,k}] ^{\text{T}}\in \mathbb{C}^{N\times 1}$ denote the output and input of the hidden layer for user $k$, $k=1,\dots,K$.

The relation between $\mathbf{D}^{\prime}$ and $\mathbf{D}$ can be expressed as
\begin{equation}\label{E:dnn}
	\overrightarrow{\mathbf{D}^{\prime}}=\sigma \big( \psi ( \overrightarrow{\mathbf{D}} )\big),
\end{equation}
where $\overrightarrow{\mathbf{D}^{\prime}}$ and $\overrightarrow{\mathbf{D}}$ denote the vectorization of $\mathbf{D}^{\prime}$ and ${\mathbf{D}}$, $\sigma\!:\!\mathbb{C}^{NK\times1}\!\rightarrow\! \mathbb{C}^{NK\times1}$ is an activation function, and $\psi\!:\!\mathbb{C}^{NK\times1}\!\rightarrow\! \mathbb{C}^{NK\times1}$ is a weight function.

The design of permutation equivariant DNNs focuses on the design of function $\psi ( \cdot )$. To highlight the novelty of the proposed UPNN, we next briefly summarize two permutation equivariant DNNs, which are the permutation equivariant neural network (PENN) designed in~\cite{9149298} and the edge-updated GNN (Edge-GNN) designed in \cite{zhao2022learning}.

\subsection{PENN}
\label{ssec:penn}
Each layer of PENN satisfies 2D-PE. For $\sigma ( \cdot )$, point-wise activation functions can be used, which do not destroy the permutation equivariance. For $\psi ( \cdot )$, PENN requires that if the input $\mathbf{D}$ is permuted as $\widetilde{\mathbf{\Pi }}\mathbf{D}{\mathbf{\Pi }}^{\text{T}}$, or equivalently, if the vectorized input $\overrightarrow{\mathbf{D}}$ is permuted as $({\mathbf{\Pi }}\otimes \widetilde{\mathbf{\Pi }})\overrightarrow{\mathbf{D}}$, then the output of $\psi ( \cdot )$ should be permuted in the same way as
\begin{align}\label{E:pe-phi}
	\psi \big( ({\mathbf{\Pi }}\otimes \widetilde{\mathbf{\Pi }})\overrightarrow{\mathbf{D}} \big) = ( {\mathbf{\Pi }}\otimes \widetilde{\mathbf{\Pi }})\psi(\overrightarrow{\mathbf{D}}),
\end{align}
where $\widetilde{\mathbf{\Pi }}\in\mathbb{R}^{N\times N}$ and $\mathbf{\Pi }\in\mathbb{R}^{K\times K}$ are two permutation matrices that change the orders of antennas and users,~respectively.

In PENN, the function $\psi(\cdot)$ has the following form
\begin{equation}\label{E:phi_W}
	\psi(\overrightarrow{\mathbf{D}})=\mathbf{W}\overrightarrow{\mathbf{D}},	
\end{equation}
where $\mathbf{W}\in \mathbb{C}^{NK\times NK}$ is the weight matrix, consisting of learnable parameters.
According to the analysis in~\cite{9149298}, $\mathbf{W}$ should have the following parameter sharing structure
\begin{align}\label{E:2D_PE_res}
	&\mathbf{W}=\left[ \begin{matrix}
		\mathbf{B}	&\mathbf{Q}&		\cdots&		\mathbf{Q}\\
		\mathbf{Q}	&\mathbf{B}&		\cdots&		\mathbf{Q}\\
		\vdots &	\vdots&		\ddots&		\vdots\\
		\mathbf{Q}&	\mathbf{Q}&		\cdots&		\mathbf{B}\\
	\end{matrix} \right],
\end{align}
where the sub-matrices $\mathbf{B},\mathbf{Q}\in \mathbb{C}^{N\times N}$ have the forms
\begin{align}\label{E:2D_PE_res1}
	\mathbf{B}=\left[ \begin{matrix}
		b&	p&		\cdots&		p\\
		p&	b&		\cdots&		p\\
		\vdots&	\vdots&		\ddots&		\vdots\\
		p&	p&		\cdots&		b\\
	\end{matrix} \right],
	\mathbf{Q}=\left[ \begin{matrix}
		q&	c&			\cdots&		c\\
		c&	q&		\cdots&		c\\
		\vdots&	\vdots&		\ddots&		\vdots\\
		c&	c&		\cdots&		q\\
	\end{matrix} \right]
\end{align}
with $ b,p,q,c\in \mathbb{C} $ representing the learnable parameters.

We can see that despite with a dimension of $NK\times NK$, $\mathbf{W}$ actually contains only four free parameters that require learning, which are independent from the values of $N$ or $K$.

\subsection{Edge-GNN}
\label{ssec:edge_gnn}
Edge-GNN was designed to learn the precoding policy over a heterogeneous graph, where BS antennas and users are defined as two types of vertices. An edge $\left( n,k\right)$ is established between antenna vertex $n$ and user vertex $k$. In Edge-GNN, the features are defined on edge, and the hidden representation of each edge $(n,k)$ is updated by the following aggregation and combination processes~\cite{zhao2022learning}.
\begin{itemize}
	\item \textbf{Aggregation:} The information aggregated at antenna vertex $n$ and user vertex $k$ is respectively
	\begin{equation}\label{E:GNN_Aggr}
		a_{n} = \sum\nolimits_{m=1,m\ne k}^K{d_{n,m}},\ \ u_{k}=\sum\nolimits_{j=1,j\ne n}^N{d_{j,k}},
	\end{equation}
	where $d_{j,k}\in \mathbb{C}$ is the hidden representation of edge $\left( j,k \right)$ in the previous layer.
	\item \textbf{Combination:}  The aggregated information is then combined with the hidden representation of edge $(n,k)$ in the previous layer as
	\begin{equation}\label{E:GNN_Comb}
		d^{\prime}_{n,k}=\sigma \left( b\cdot d_{n,k}+p\cdot u_{k}+q\cdot a_{n}\right) ,
	\end{equation}
	where $\sigma ( \cdot )$ is a point-wise activation function and $ b,p,q\in \mathbb{C} $ are learnable parameters.
\end{itemize}

By expressing the update process in a matrix form similar to \eqref{E:phi_W}, the weight matrix $\mathbf{W}$ has the same parameter sharing structure as in \eqref{E:2D_PE_res},
where the sub-matrices $\mathbf{B},\mathbf{Q}\in \mathbb{C}^{N\times N}$ have the forms as in \eqref{E:2D_PE_res1} but $c=0$.

It is apparent that Edge-GNN uses one fewer learnable parameters in $\mathbf{W}$ than PENN. This reduction leads to enhanced learning performance and a decrease of training complexity, as to be shown later.

\section{Design of UPNN}
\label{sec:pagestyle}

In this section, we derive the parameter sharing structure of the weight matrix that satisfies the UE-PE property and demonstrate its shortcomings in learning the optimal precoder. We then develop a novel non-linear weighting process, based on which we propose a joint unitary and permutation equivariant DNN.


\subsection{Structure of Weight Matrix with the UE-PE Property}
\label{ssec:upnn_tra_W}
Let us first consider that the weight function $\psi(\cdot)$ has the form $\psi\big(\overrightarrow{\mathbf{D}}\big)=\mathbf{W}\overrightarrow{\mathbf{D}}$ as the PENN in \eqref{E:phi_W}, where the weight matrix $\mathbf{W}$ consists of learnable parameters.

To achieve the UE-PE property, $\psi(\cdot)$ needs to satisfy
\begin{align}\label{E:UPNN-phi}
	\psi \big( ({\mathbf{\Pi }}\otimes {\mathbf{U}})\overrightarrow{\mathbf{D}} \big) = ( {\mathbf{\Pi }}\otimes {\mathbf{U}})\psi(\overrightarrow{\mathbf{D}}).
\end{align}

With $\psi(\overrightarrow{\mathbf{D}})=\mathbf{W}\overrightarrow{\mathbf{D}}$, \eqref{E:UPNN-phi} can be rewritten as
\begin{equation}\label{E:upnn_tra_E_12}
	\left( \mathbf{\Pi }\otimes \mathbf{U} \right) \mathbf{W}\overrightarrow{\mathbf{D}}=\mathbf{W}\left( \mathbf{\Pi }\otimes \mathbf{U} \right)\overrightarrow{\mathbf{D}}.
\end{equation}

To ensure that \eqref{E:upnn_tra_E_12} holds for arbitrary $\overrightarrow{\mathbf{D}}$, it can be readily proved that the following condition must hold
\begin{equation}\label{E:upnn_tra_E_1}
	\left( \mathbf{\Pi }\otimes \mathbf{U} \right) \mathbf{W}=\mathbf{W}\left( \mathbf{\Pi }\otimes \mathbf{U} \right),
\end{equation}
which requires $\mathbf{W}$ to have a particular parameter sharing structure.

To find the structure of $\mathbf{W}$ satisfying \eqref{E:upnn_tra_E_1}, we first investigate the structures of the two matrices $\mathbf{W}_S\in\mathbb{C}^{K\times K}$ and $\mathbf{W}_U\in\mathbb{C}^{N\times N}$ that satisfy
\begin{subequations} \label{E:pe_con}
	\begin{align}
		\mathbf{\Pi W}_S=\mathbf{W}_S\mathbf{\Pi}, \label{E:pe_cona}\\
		\mathbf{UW}_U=\mathbf{W}_U\mathbf{U}. \label{E:pe_conb}
	\end{align}
\end{subequations}

According to the analysis in~\cite{9149298}, we know that $\mathbf{W}_S$ has the following parameter sharing structure
\begin{equation}\label{E:pe}
	\mathbf{W}_S=\left[ \begin{matrix}
		b_s&	q_s&		\cdots&		q_s\\
		q_s&	b_s&		\cdots&		q_s\\
		\vdots&	\vdots&		\ddots&		\vdots\\
		q_s&	q_s&		\cdots&		b_s\\
	\end{matrix} \right].
\end{equation}

The permutation matrix is a special unitary matrix, hence the structure in \eqref{E:pe} is a necessary condition for $\mathbf{W}_U$, i.e.,  $[\mathbf{W}_U]_{ii}$ can be defined as $[\mathbf{W}_U]_{ii} = b_u$ and $[\mathbf{W}_U]_{ij} = q_u$, $i\neq j$. Upon substituting this $\mathbf{W}_U$ into \eqref{E:pe_conb}, we obtain an equality for the elements on the first row and first column of both sides as
\begin{equation}\label{E:ue_con1}
	b_u[\mathbf{U}]_{1,1}  + q_u\sum\nolimits_{j=2}^N [\mathbf{U}]_{1,j} = b_u[\mathbf{U}]_{1,1} + q_u\sum\nolimits_{i=2}^N [\mathbf{U}]_{i,1}.
\end{equation}
Condition \eqref{E:ue_con1} holds for all $\mathbf{U}$, leading to that $q_u=0$ must hold. Then, $\mathbf{W}_U$ has the following structure
\begin{equation}\label{E:ue}
	\mathbf{W}_U=\left[ \begin{matrix}
		b_u&	0&		\cdots&		0\\
		0&	b_u&		\cdots&		0\\
		\vdots&	\vdots&		\ddots&		\vdots\\
		0&	0&		\cdots&		b_u\\
	\end{matrix} \right].
\end{equation}


Further applying the property of Kronecker product to \eqref{E:pe_cona} and \eqref{E:pe_conb}, we obtain
\begin{equation}\label{E:ue_pe_con_2}
	\left( \mathbf{\Pi }\otimes \mathbf{U} \right) \left( \mathbf{W}_S\otimes \mathbf{W}_U \right) =\left( \mathbf{W}_S\otimes \mathbf{W}_U \right) \left( \mathbf{\Pi }\otimes \mathbf{U} \right).
\end{equation}

By comparing \eqref{E:upnn_tra_E_1} and \eqref{E:ue_pe_con_2}, we can find that the weight matrix $\mathbf{W}$ equals to $\mathbf{W}_S\otimes \mathbf{W}_U$, which can be expressed as
\begin{equation}\label{E:pe2}
	\mathbf{W}=\left[ \begin{matrix}
		b_sb_u&	q_sb_u&		\cdots&		q_sb_u\\
		q_sb_u&	b_sb_u&		\cdots&		q_sb_u\\
		\vdots&	\vdots&		\ddots&		\vdots\\
		q_sb_u&	q_sb_u&		\cdots&		b_sb_u\\
	\end{matrix} \right]\otimes \mathbf{I}_N.
\end{equation}

By defining  $b=b_sb_u$ and $q= q_sb_u$, we finally obtain that the weight matrix $\mathbf{W}$ satisfying condition \eqref{E:upnn_tra_E_1} have the following parameter sharing structure
\begin{equation}\label{E:W_un_pi_new}
	\mathbf{W}=\left[ \begin{matrix}
		b&	q&		\cdots&		q\\
		q&	b&		\cdots&		q\\
		\vdots&	\vdots&		\ddots&		\vdots\\
		q&	q&		\cdots&		b\\
	\end{matrix} \right] \otimes \mathbf{I}_N \triangleq \mathbf{\Omega}\otimes \mathbf{I}_N,
\end{equation}
where $b, q\in\mathbb{C}$ are two learnable scalar parameters.

\begin{figure}
	\centering
	\includegraphics[width=0.44\textwidth]{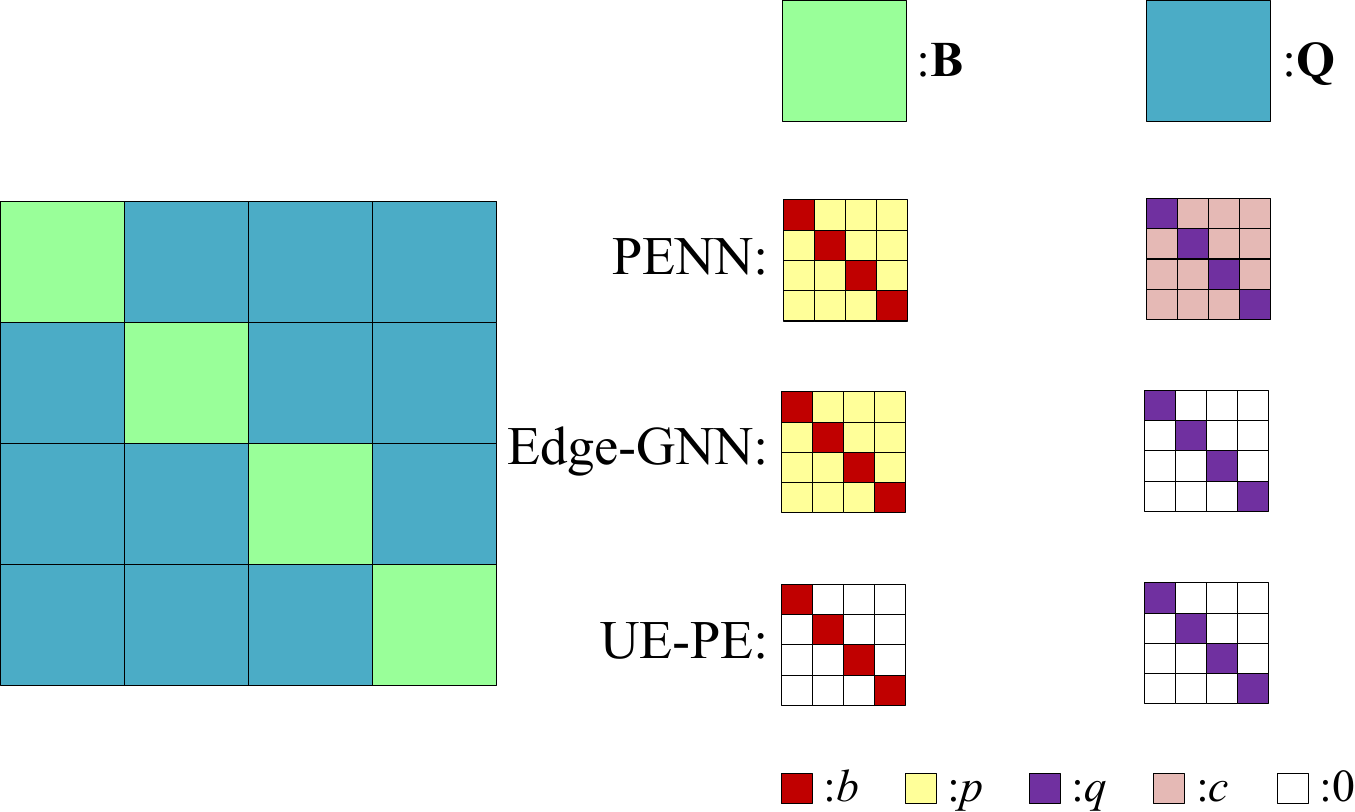}
	\captionsetup{font=small, labelsep=period}
	\caption{Parameter sharing structures for weight matrices of PENN, Edge-GNN, and that satisfying UE-PE.}
	\label{F:Com_NNs}
\end{figure}

Fig.~\ref{F:Com_NNs} illustrates the parameter sharing structures for weight matrices $\mathbf{W}$ of PENN, Edge-GNN, and $\mathbf{W}$ satisfying UE-PE. On the left side, each sub-block represents a sub-matrix, and the sub-blocks with the same color indicate the same sub-matrices. On the right side, the two columns show the sub-matrices $\mathbf{B}$ and $\mathbf{Q}$, respectively. The detailed structures of $\mathbf{B}$ and $\mathbf{Q}$ are also illustrated for three conditions. It demonstrates that Edge-GNN eliminates one learnable parameter $c$ in $\mathbf{Q}$ compared to PENN. UE-PE further reduces an additional learnable parameter $p$ in $\mathbf{B}$ compared to Edge-GNN.


\subsection{Loss of Learning Capability}
\label{ssec:lea_loss}
With \eqref{E:W_un_pi_new}, the output of the weight function $\psi ( \overrightarrow{\mathbf{D}})$, denoted by $\overrightarrow{\mathbf{D}''}$, can be expressed as
\begin{align} \label{E:tildeD}
	\overrightarrow{\mathbf{D}''} \triangleq \psi ( \overrightarrow{\mathbf{D}} ) =(\mathbf{\Omega}\otimes \mathbf{I}_N) \overrightarrow{\mathbf{D}}
	=\overrightarrow{\mathbf{I}_N\mathbf{D}\mathbf{\Omega}^{\text{T}}},
\end{align}
where $\mathbf{D}''=[\mathbf{d}''_{1}, \dots, \mathbf{d}''_{K}]$ with $\mathbf{d}''_{k}=[ d''_{1,k}, \dots, d''_{N,k}] ^{\text{T}}\in \mathbb{C}^{N\times 1}$ denoting the output of $\psi ( \overrightarrow{\mathbf{D}} )$ for user $k$, and the final equality follows from the property of Kronecker product. We can obtain that $\mathbf{D}'' = \mathbf{D}\mathbf{\Omega}^{\text{T}}$. Thus, the $k$-th column of $\mathbf{D}''$ equals to the product of $\mathbf{D}$ by the $k$-th column of $\mathbf{\Omega}^{\text{T}}$, which can be expressed as
\begin{equation}\label{E:tra_update}
	\mathbf{d}''_{k} = b{\mathbf{d}}_{k}+\sum\nolimits_{m=1,m\ne k}^K{q{\mathbf{d}}_{m}},
\end{equation}
where $b$ and $q$ are two learnable scalar parameters in $\mathbf{\Omega}$ as defined in \eqref{E:W_un_pi_new}.

It is shown from \eqref{E:tra_update} that the $n$-th element of $\mathbf{d}''_{k}$ only depends on the $ n $-th element of $ \mathbf{d}_{m}$, $m=1,\cdots K$. As a result, the update of the $n$-th row of $\mathbf{D}''$ relies solely on the $n$-th row of $\mathbf{D}$. This indicates that the weight function $\psi(\cdot)$ is in fact a point-wise function. If we also employ a poise-wise activation function $\sigma(\cdot)$ at each layer, the $n$-th row of the resulting precoder $\mathbf{V}$ only depends on the $n$-th row of the input channel $\mathbf{H}$. This is obviously far away from the structure of the optimal precoder, typically represented as $(\mathbf{H}\mathbf{H}^{\text{H}}+\nu\mathbf{I}_N)^{-1}\mathbf{H}$~\cite{6832894}.

The diminished capability of learning the optimal precoder indicates a mismatch between the policy being learned and the functions that can be learned by the DNN with the derived weight matrix in \eqref{E:W_un_pi_new}. Hence, while DNNs with parameter sharing structure are effective for learning policies satisfying permutation equivariance, they fall short for the stronger unitary equivariance property.

\subsection{UPNN with Non-linear Weighting Process}
\label{ssec:scalar}
Given the shortcoming of the weight function $\psi\big(\overrightarrow{\mathbf{D}}\big)=\mathbf{W}\overrightarrow{\mathbf{D}}$ with parameter sharing structure, we next construct a non-linear weight function $\psi\big(\overrightarrow{\mathbf{D}}\big)$, which has the following structure
\begin{align}\label{E:gen_W}
	\psi ( \overrightarrow{\mathbf{D}} ) &=\left(\left[ \begin{matrix}
			b\mathbf{d}_{1}^{ \text{H}}\mathbf{d}_{1}&	q\mathbf{d}_{1}^{ \text{H}}\mathbf{d}_{2}&		\cdots&		q\mathbf{d}_{1}^{ \text{H}}\mathbf{d}_{K}\\
			q\mathbf{d}_{2}^{ \text{H}}\mathbf{d}_{1}&	b\mathbf{d}_{2}^{ \text{H}}\mathbf{d}_{2}&		\cdots&		q\mathbf{d}_{2}^{ \text{H}}\mathbf{d}_{K}\\
			\vdots&	\vdots&		\ddots&		\vdots\\
			q\mathbf{d}_{K}^{ \text{H}}\mathbf{d}_{1}&	q\mathbf{d}_{K}^{ \text{H}}\mathbf{d}_{2}&		\cdots&		b\mathbf{d}_{K}^{ \text{H}}\mathbf{d}_{K}\\
		\end{matrix} \right]\otimes\mathbf{I}_N \right) \overrightarrow{\mathbf{D}}\nonumber\\
& \triangleq \left(\mathcal{G}\left( \mathbf{D} \right)\otimes\mathbf{I}_N\right)\overrightarrow{\mathbf{D}},
\end{align}
where the diagonal elements of $\mathcal{G}\left( \mathbf{D} \right)$ are $ b \mathbf{d}_{k}^{\text{H}}\mathbf{d}_{k}$, the off-diagonal elements are $q \mathbf{d}_{k}^{\text{H}}\mathbf{d}_{m}$ for $k\neq m$, and $b, q\in \mathbb{C}$ are two learnable scalar parameters. We can note that each term in the weight matrix not only contains learnable parameters, but also is a non-linear function about the input of the hidden layer. The UPNN we designed exactly uses the structure in \eqref{E:gen_W} as the weight function $\psi\big(\overrightarrow{\mathbf{D}}\big)$ in \eqref{E:dnn}.

We next verify that the constructed non-linear weight function $\psi ( \overrightarrow{\mathbf{D}} )$ satisfies the UE-PE condition given by \eqref{E:UPNN-phi}. By substituting \eqref{E:gen_W} into \eqref{E:UPNN-phi}, we just need to verify equivalently that $\mathcal{G}\left( \mathbf{D} \right)$ satisfies
\begin{align}\label{E:equi_con}
	\left( \mathcal{G}\left( \mathbf{UD\Pi }^{\text{T}} \right) \mathbf{\Pi }\otimes \mathbf{U} \right) \overrightarrow{\mathbf{D}}
	=\left( \mathbf{\Pi} \mathcal{G}\left( \mathbf{D} \right) \otimes \mathbf{U} \right) \overrightarrow{\mathbf{D}}.
\end{align}


First, it is evident that the elements of $\mathcal{G}\left( \mathbf{D} \right)$ are unitary invariant scalar functions, i.e., $\left( \mathbf{Ud}_k \right) ^{\text{H}}\left( \mathbf{Ud}_m \right) =\mathbf{d}_{k}^{\text{H}}\mathbf{d}_m$. Hence, $\mathcal{G}\left( \mathbf{D} \right)$ satisfies
\begin{align}\label{E:G_con_1}
	\mathcal{G}\left( \mathbf{UD} \right) &=\mathcal{G}\left( \mathbf{D} \right).
\end{align}

We now confirm that $\mathcal{G}\left( \mathbf{D} \right)$ satisfies
\begin{align}\label{E:G_con_2}
	\mathcal{G}\left( \mathbf{D\Pi }^{\text{T}} \right) \mathbf{\Pi }&=\mathbf{\Pi }\mathcal{G}\left( \mathbf{D} \right).
\end{align}
Note that any permutation matrix $\mathbf{\Pi}$ can be decomposed into a sequence of elementary permutation matrices, where an elementary permutation matrix involves the interchange of two adjacent rows or columns. Thus, without loss of generality, we consider the permutation matrix $\mathbf{\Pi}_{12}$ that interchanges the first and second rows or columns. Then, we can obtain
\begin{equation}\label{E:gen_UPE_equ1}
	\mathcal{G}\left( \mathbf{D} \mathbf{\Pi}_{12}^{\text{T}}\right)  =\left[ \begin{matrix}
		b\mathbf{d}_{2}^{ \text{H}}\mathbf{d}_{2}&	q\mathbf{d}_{2}^{ \text{H}}\mathbf{d}_{1}&		\cdots&		q\mathbf{d}_{2}^{ \text{H}}\mathbf{d}_{K}\\
		q\mathbf{d}_{1}^{ \text{H}}\mathbf{d}_{2}&	b\mathbf{d}_{1}^{ \text{H}}\mathbf{d}_{1}&		\cdots&		q\mathbf{d}_{1}^{ \text{H}}\mathbf{d}_{K}\\
		\vdots&	\vdots&		\ddots&		\vdots\\
		q\mathbf{d}_{K}^{ \text{H}}\mathbf{d}_{2}&	b\mathbf{d}_{K}^{ \text{H}}\mathbf{d}_{1}&		\cdots&		q\mathbf{d}_{K}^{ \text{H}}\mathbf{d}_{K}\\
	\end{matrix} \right],
\end{equation}
\begin{equation}\label{E:gen_UPE_equ2}
	\mathcal{G}\left( \mathbf{D} \mathbf{\Pi}_{12}^{\text{T}}\right)\mathbf{\Pi}_{12}  =\left[ \begin{matrix}
		q\mathbf{d}_{2}^{ \text{H}}\mathbf{d}_{1} &	b\mathbf{d}_{2}^{ \text{H}}\mathbf{d}_{2}&		\cdots&		q\mathbf{d}_{2}^{ \text{H}}\mathbf{d}_{K}\\
		b\mathbf{d}_{1}^{ \text{H}}\mathbf{d}_{1}&	q\mathbf{d}_{1}^{ \text{H}}\mathbf{d}_{2}&		\cdots&		q\mathbf{d}_{1}^{ \text{H}}\mathbf{d}_{K}\\
		\vdots&	\vdots&		\ddots&		\vdots\\
		q\mathbf{d}_{K}^{ \text{H}}\mathbf{d}_{1}&q\mathbf{d}_{K}^{ \text{H}}\mathbf{d}_{2}	&		\cdots&		b\mathbf{d}_{K}^{ \text{H}}\mathbf{d}_{K}\\
	\end{matrix} \right].
\end{equation}
We can find that $\mathcal{G}\left( \mathbf{D} \mathbf{\Pi}_{12}^{\text{T}}\right)\mathbf{\Pi}_{12} $ exactly equals to the interchange of the first and second row of $\mathcal{G}\big( \mathbf{D} \big)$, i.e., $ \mathbf{\Pi}_{12}\mathcal{G}\left( \mathbf{D}\right)$. It indicates that $\mathcal{G}\left( \mathbf{D} \right)$ satisfies the condition in \eqref{E:G_con_2}.

By substituting \eqref{E:G_con_1} and \eqref{E:G_con_2} into the left-hand side of \eqref{E:equi_con}, we can obtain
\begin{align}\label{E:equi_con_1}
	\left( \mathcal{G}\left( \mathbf{UD\Pi }^{\text{T}} \right) \mathbf{\Pi }\otimes \mathbf{U} \right) \overrightarrow{\mathbf{D}}&=\left( \mathcal{G}\left( \mathbf{D\Pi }^{\text{T}} \right) \mathbf{\Pi }\otimes \mathbf{U} \right) \overrightarrow{\mathbf{D}}\\
	&=\left(\mathbf{\Pi }\mathcal{G}\left( \mathbf{D} \right)\otimes \mathbf{U} \right)\overrightarrow{\mathbf{D}}\nonumber,
\end{align}
where the first equality follows from \eqref{E:G_con_1}, and the second equality follows from \eqref{E:G_con_2}. Hence, we finally verified that the UE-PE condition \eqref{E:equi_con} is satisfied, and the resulting DNN based on the non-linear weighting process in \eqref{E:gen_W} is called \emph{UPNN}.

Furthermore, the function $\psi ( \overrightarrow{\mathbf{D}} )$ can be expressed as consisting of aggregation and combination processes similar to those in Edge-GNN. Specifically, the $k$-th sub-vector of the output of $\psi(\cdot)$ can be expressed as
\begin{align}\label{E:gen_UPE_update}
	\mathbf{d}''_{k}&= b\big(\mathbf{d}_{k}^{ \text{H}}\mathbf{d}_{k} \big) \mathbf{d}_{k} +\sum\nolimits_{m=1,m\ne k}^K{q\big(\mathbf{d}_{k}^{ \text{H}}\mathbf{d}_{m} \big) \mathbf{d}_{m}},
\end{align}
where the aggregation and combination processes can be obtained as follows.
\begin{itemize}
	\item \textbf{Aggregation:}
	\begin{equation}\label{E:UPNN_Aggr}
		a_{n}=\sum\nolimits_{m=1,m\ne k}^K{\big( \mathbf{d}_{k}^{ \text{H}}\mathbf{d}_m \big) d_{n,m}}.
	\end{equation}
	\item \textbf{Combination:}
	\begin{equation}\label{E:UPNN_Comb}
		d''_{n,k}=b\big( \mathbf{d}_{k}^{ \text{H}}\mathbf{d}_k \big) d_{n,k}+q\cdot a_{n}.
	\end{equation}
\end{itemize}

The aggregation and combination of UPNN exhibit two key differences from those of the Edge-GNN. First, the UPNN only aggregates information from the edges connected to adjacent user vertices. Therefore, UPNN requires one fewer learnable parameter than Edge-GNN at each layer. Second, in UPNN, the aggregation and combination are weighted by the inner product of the input vectors $ \mathbf{d}_{m}$. As a result, the output $d''_{n,k}$ in \eqref{E:UPNN_Comb} is not merely dependent on the $n$-th element of $\mathbf{d}_{m}$. This effectively addresses the issue induced by the weight function in Section \ref{ssec:lea_loss}.

\section{Simulation Results}
\label{sec:results}

In this section, we evaluate the performance and learning efficiency of the proposed UPNN and compare it with the following methods.
\begin{itemize}
	\item \textbf{WMMSE}: This is a widely used numerical algorithm for solving the sum rate maximization problem~\cite{shi2011iteratively}, which can obtain at least a local optimal solution.	
	\item \textbf{FNN}: This is a standard fully connected DNN.	
	\item \textbf{Edge-GNN}: This is the GNN designed in \cite{zhao2022learning} leveraging both permutation prior and topology prior.
	\item \textbf{PENN}: This is the DNN designed in \cite{9149298} leveraging permutation equivariance.
	\item \textbf{Edge-GAT}: This is the edge-updated GAT designed in \cite{10333362} leveraging attention mechanism.
	\item \textbf{NormNN}: This is the unitary equivariant DNN designed in~\cite{ma2023feedforward}.
	\item \textbf{Model-GNN}: This is the GNN proposed in \cite{guo2023model}, which leverages both permutation equivariance and the model of Taylor's expansion of matrix pseudo-inverse.
\end{itemize}

In simulations, we use the sum rate achieved by WMMSE as a benchmark to normalize the sum rate of the learning methods. For WMMSE, we consider 50 random initial values and then choose the best result to represent its performance. All learning methods are trained by unsupervised learning, employing the negative sum rate as the loss function, and the power constraint can be satisfied by passing the output of the DNNs through a normalization layer, i.e., $\frac{\sqrt{P_{\text{max}}}\mathbf{V}}{\lVert \mathbf{V} \rVert _F}$. The learning methods are well trained via fine tuning the hyperparameters, which are shown in Table~\ref{tab0}.  To mitigate the impact of randomness from sample selection and initial weight settings, we independently train five different DNNs and use average test performance for evaluation. Each trained DNN is tested on 2000 samples. The channel samples are generated based on Rayleigh distribution, and the SNR is 10~dB. The simulation results are obtained on a computer with a 6-core Intel i7-8700K CPU, an NVIDIA GeForce GTX 1080Ti GPU.

\begin{table}
	\renewcommand{\arraystretch}{1.4}
	\centering
	\caption{Hyper-parameters} \label{tab0}
	\footnotesize
	\begin{tabular}{c|c|c|c}
		\hline\hline
		& \makecell[c]{Hidden\\layers} & \makecell[c]{Hidden\\representations} & \makecell[c]{Learning\\rate}  \\
		\hline
		\bf FNN & 6 & $[128]\times6$ & $ 4\times10^{-4} $  \\
		\hline
		\bf Edge-GNN & 5 & $[128]\times4 + [32]$ 		
		* & $ 4\times10^{-4} $   \\
		\hline
		\bf PENN & 5 & $[128]\times4 + [32]$ & $ 4\times10^{-4} $   \\
		\hline
		\bf Edge-GAT & 5 & $[128]\times4 + [32]$ & $ 4\times10^{-4} $   \\
		\hline
		\bf NormNN & 5 & $[128]\times4 + [32]$ & $ 4\times10^{-4} $  \\
		\hline
		\bf Model-GNN & 4 & $[16]\times3 + [4]$ & $ 1\times10^{-2} $ \\
		\hline
		\bf UPNN & 4 & $[16]\times3 + [4]$ &$ 1\times10^{-2} $ \\
		\hline\hline
	\end{tabular}
	\begin{flushleft}
		{\footnotesize 		
			*: $[128]\times4 + [32]$ means that the first 4 hidden layers use 128 hidden representations and the final hidden layer uses 32 hidden representations.}
	\end{flushleft}
\end{table}

\begin{figure}
	\centering
	\includegraphics[width=0.47\textwidth]{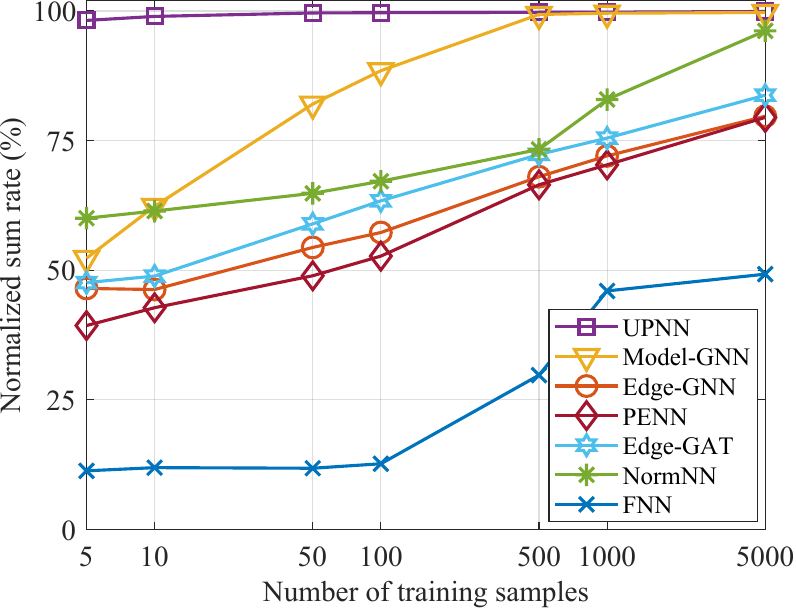}
	\captionsetup{font=small, labelsep=period}
	\caption{Learning performance vs the number of training samples with $N=8$ and $K=4$.}
	\label{F:7}
\end{figure}

\label{ssec:performance}
In Fig.~\ref{F:7}, the DNNs are trained with eight BS antennas and four users. By comparing Edge-GNN with PENN, we can observe the gain of Edge-GNN owing to fewer learnable parameters. Edge-GAT performs better than Edge-GNN because it leverages the attention mechanism. UPNN exhibits evident gain over NormNN. When trained with a smaller set of samples, UPNN substantially outperforms other learning~methods.

%

%

\begin{table*}
	\footnotesize 
	\renewcommand{\arraystretch}{1.4}
	\centering
	\caption{Training complexity and inference time} \label{tab1}
	\begin{tabular}{c|c|c|c|c|c|c|c|c}
		\hline\hline
		&& \bf FNN & \bf Edge-GNN& \bf PENN & \bf Edge-GAT & \bf NormNN & \bf Model-GNN & \bf UPNN \\
		\hline
		\multirow{4}{*}{\makecell{$N=8$ \\ $K=4$}}
		& Parameter & 321.14 M & 0.61 M & 0.81 M & 0.62 M & 0.41 M & 14.17 K & 5.52 K \\
		\cline{2-9}
		& Sample & $>$200,000* & 180,000 & $>$200,000* & 150,000  & 10,000 & 300 & 15  \\
		\cline{2-9}
		& Training time & $>$6 h 37 min & 5 min 44 s & 30 min 20 s & 22 min 58 s & 13 min 15 s & 5.00 s & 0.75 s  \\
		\cline{2-9}
		& Inference time & 4.29 ms & 1.59 ms & 6.52 ms & 3.37 ms & 3.66 ms & 3.58 ms & 2.81 ms  \\
		\hline
		\multirow{4}{*}{\makecell{$N=16$ \\ $K=8$}}
		& Parameter & 324.14 M & 0.61 M & 0.81 M & 0.62 M & 0.41 M & 14.17 K & 5.52 K \\
		\cline{2-9}
		& Sample & $>$200,000* & $>$200,000* & $>$200,000* & $>$200,000*  &$>$200,000* & 70 & 5 \\
		\cline{2-9}
		& Training time & $>$6 h 44 min & $>$1 h 53 min & $>$2 h 33 min & $>$5 h 43 min & $>$4 h 18 min & 7.82 s & 1.09 s \\
		\cline{2-9}
		& Inference time & 4.34 ms & 1.74 ms & 12.02 ms & 3.39 ms & 3.85 ms & 3.66 ms & 2.87 ms \\
		\hline\hline
	\end{tabular}
	\begin{flushleft}
		{\footnotesize 		
			*: For the cases marked by *, the 98\% performance cannot be achieved no matter how many training samples are used. Thus, we record the training time and inference time with 200 epochs and 200,000 training samples. }
	\end{flushleft}
\vspace{-0.2cm}
\end{table*}

\begin{figure}
	\centering
	\includegraphics[width=0.47\textwidth]{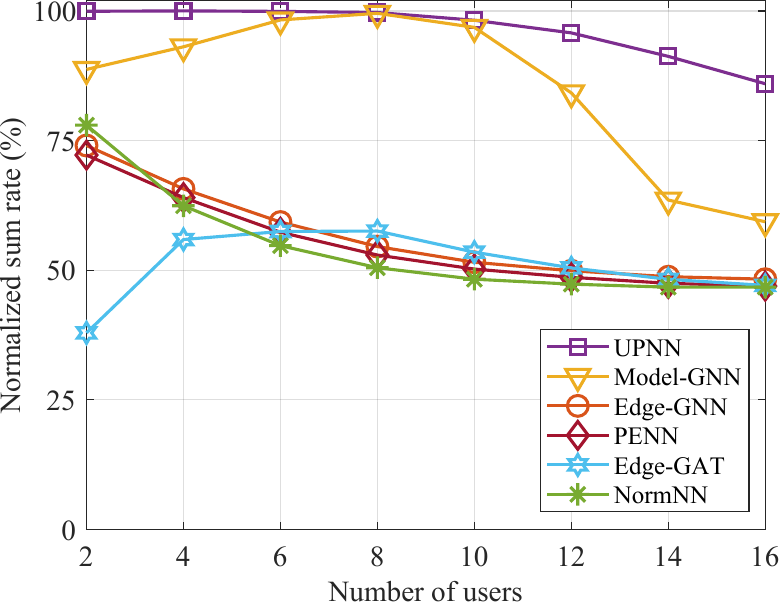}
	\captionsetup{font=small, labelsep=period}
	\caption{Generalization performance to the number of users with $N=16$ and 500 training samples.}
	\label{F:10}
\end{figure}

In Fig.~\ref{F:10}, we compare the generalization performance of the learning methods to the number of users. The comparison considers five equivariant DNNs, excluding FNN due to its lack of generalizability. The DNNs are trained in a scenario with $ K=8$, and are tested in various scenarios where the number of users ranges from 2 to 16. It can be observed that the performance of Edge-GNN, PENN and NormNN degrades as the number of test users increases. Model-GNN and Edge-GAT exhibit decrease in performance when the number of test users differs from the number of training users. UPNN  exhibits superior generalizability. While its performance also degrades with a larger number of test users, this degradation is less pronounced compared to other learning~methods.

\begin{figure}
	\centering
	\includegraphics[width=0.47\textwidth]{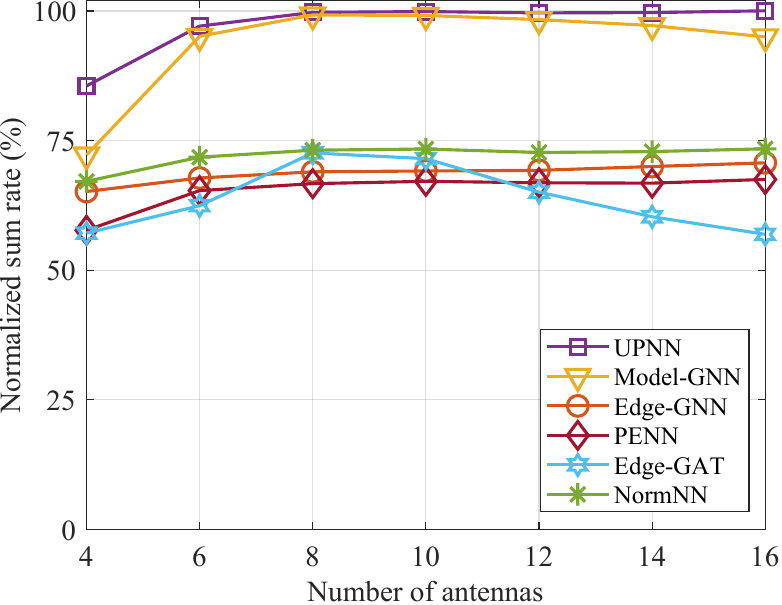}
	\captionsetup{font=small, labelsep=period}
	\caption{Generalization performance to the number of antennas with $K=4$ and 500 training samples.}
	\label{F:11}
\end{figure}

In Fig.~\ref{F:11}, we show the generalization performance of the learning method to the number of antennas. The DNNs are trained with $ N=8 $ and then tested in scenarios with varying numbers of antennas ranging from 4 to 16. We can find that the generalization performance of Model-GNN exhibits a large degradation when the number of test antennas is small. In this case, UPNN also experiences a performance degradation, but the extent is much less severe.

\label{ssec:complexity}
In Table~\ref{tab1}, we compare training complexity of different learning methods. Specifically, we examine the number of training samples and the number of learnable parameters (represented in million as `M' and thousand as `K') required by different DNNs to achieve the same normalized sum rate (set as 98\%). We also compare the training time and inference time on GPU. The inference time is obtained as the average running time over 2000 test samples. The results show that UPNN demands the least training complexity.

\section{Conclusions}
\label{sec:majhead}

In this paper, we proposed a novel joint unitary and permutation equivariant neural network (UPNN) for precoder learning. We first shown that equivariant DNNs with weight matrix parameter sharing structure derived from permutation equivariance property can well learn the precoding policy, but those derived from unitary equivariance property are unable to learn the optimal precoding policy. Then, we developed a unitary and permutation equivariant non-linear weighting process, based on which the UPNN was proposed. Simulation results showed that the proposed UPNN outperforms baseline equivariant DNNs in terms of learning performance, generalization performance, and training complexity.


\bibliographystyle{IEEEtran}
\bibliography{UPNN}

\end{document}